\documentstyle[12pt,aaspp4]{article}
\lefthead{Ogle et al.}
\righthead{Scattered Broad H$\alpha$ in Cygnus A}
\slugcomment{ApJ Letters, accepted 3/20/97}

\begin{document}

\title{Scattered Nuclear Continuum and Broad H$\alpha$ in Cygnus A}

\author{P. M. Ogle, M. H. Cohen}
\affil{Mail Stop 105-24, California Institute of Technology, Pasadena, 
       CA 91125}
\authoraddr{Electronic mail: pmo@astro.caltech.edu}
\author{J. S. Miller, H. D. Tran \altaffilmark{1}}
\affil{UCO/Lick Observatory, University of California, Santa Cruz, 
       CA 95064}
\altaffiltext{1}{Current address: Institute of Geophysics and Planetary 
       Physics, Lawrence Livermore National Laboratory L-413, Livermore, CA 
       94550}
\author{R. A. E. Fosbury \altaffilmark{2}} 
\affil{Space Telescope--European Coordinating Facility, Karl-Schwarzschild 
       Str. 2, D-85748, Garching bei Munchen, Germany}
\altaffiltext{2}{Affiliated to the Astrophysics Division, Space Science 
       Department, European Space Agency.}
\and
\author{R. W. Goodrich}
\affil{The W.M. Keck Observatory, 65-1120 Mamalahoa Highway, Kamuela, HI
       96743}

\begin{abstract}

We have discovered scattered broad Balmer emission lines in the spectrum of 
Cygnus A, using the Keck II telescope. Broad H$\alpha$ appears in polarized 
flux from components on either side of the nucleus, and to a lesser extent in 
the nucleus. The full-width at half-maximum of broad H$\alpha$ is 26,000 km 
s$^{-1}$, comparable to the widest emission lines seen in broad-line radio 
galaxies. Scattered AGN light provides a significant contribution to the total
flux  at 3800~\AA ~(rest) of the western component, where the polarization 
rises to 16\%. The spatially integrated flux of Cygnus A at 5500~\AA ~can
be decomposed into an elliptical galaxy fraction ($F_g$=0.70), a highly
polarized blue component (FC1=0.15), a less polarized red component (FC=0.09),
and a contribution from the nebular continuum (0.06). Imaging polarimetry 
shows a double fan of polarization vectors with circular symmetry which 
corresponds to the ionization cone seen in HST images. Our results are 
consistent with scattering of light from a hidden quasar of modest luminosity 
by an extended, dusty narrow-line region.

\end{abstract}

\keywords{galaxies: active --- galaxies: individual (Cygnus A)}

\section{Introduction}

In recent years there has been considerable interest in the unification
of active galaxies. In part this was stimulated by the discovery 
(\cite{am85}) that the nucleus of the Seyfert 2 galaxy NGC 1068 contains a 
broad-line region (BLR) obscured from direct view, but visible in scattered 
light. This result gives support to the hypothesis that Seyfert type 1 and 
type 2 galaxies are the same kind of object seen from different aspects. 
Barthel (1989) extends this picture to quasars and radio galaxies, proposing 
that narrow-line radio galaxies (NLRG) of Fanaroff-Riley type 2 contain 
quasars that are obscured from our viewing direction.  He finds that
this hypothesis is consistent with the relative numbers of quasars and 
appropriate radio galaxies in the 3C catalogue if the quasar is visible from 
any direction within a cone of half-opening angle about $45 \arcdeg$.  Beyond 
that cone the optical continuum source and BLR would be obscured from direct 
view. 

Attempts to detect hidden BLR in radio galaxies have had some success
(e.g., 3C 234, \cite{a84} and \cite{tcg95}; 3C 321, \cite{yet96}). High 
polarization perpendicular to the radio axis and broad Mg II suggest the 
presence of hidden quasars in high redshift radio galaxies as well
(e.g., \cite{scf94}; \cite{cet96}). According to the Barthel hypothesis, the 
prototypical NLRG containing the exceptionally strong radio source Cygnus A 
(3C 405) should harbor an obscured quasar. Djorgovski et al. (1991) present 
evidence from infrared observations that Cygnus A does contain a luminous 
quasar.  Imaging polarimetry by Tadhunter, Scarrott, \& Rolph (1990) suggests 
that there is an extended region which scatters light from a hidden nucleus. 
Broad Mg II is detected by Antonucci, Hurt, \& Kinney (AHK, 1994) in the  
ultraviolet. In spite of considerable effort (e.g., \cite{gm89}; \cite{jt93}),
previous attempts to detect broad lines in polarized flux have been
unsuccessful. In this Letter we present spectropolarimetry from the Keck II 
Telescope that shows broad lines seen in scattered light originating from an 
obscured nuclear region. Cygnus A does indeed contain an obscured AGN with 
broad permitted lines, giving additional weight to the radio galaxy--quasar 
unification hypothesis. Section 2 summarizes the observations, and \S 3 and 
\S 4 present the results and their implications, respectively.

\section{Observations and Reductions}

We observed Cygnus A with the LRIS polarimeter on 1996 October 4 and 5 
with the Keck II Telescope (four nights after scientific commissioning). The 
first night of observations included one hour of $B$ band imaging polarimetry.
On the second night 2.2 hours of spectropolarimetry were taken with a 300 line
mm$^{-1}$ grating blazed at 5000~\AA.  A $1\arcsec$ slit was placed at a 
position angle of $101\arcdeg$ to include the highly polarized western knot, 
the eastern knot, and the nuclear knot. This slit location (indicated in Fig. 
1) is nearly along the radio axis, which is at $104\arcdeg$ on kpc scales.
The data are reduced using standard procedures (\cite{mrg88}; \cite{cet97}). 
Prior to calculating the Stokes parameters ($Q$, $U$), the flux was binned by 
4 pixels to improve the signal-to-noise ratio (SNR), which drops below 3.0 for 
$Q$ at 7200~\AA~ (rest). The fractional polarization ($P$) is estimated by 
rotating the Stokes parameters. The spectroscopy is calibrated using 
BD $+28\arcdeg ~4211$ as a flux standard and HD 155528 as a polarimetric 
standard. On the first night the telescope would not come to a sharp focus so 
stellar images are elongated NW-SE with a seeing ellipse of $1.0\arcsec \times 
0.8\arcsec$. The seeing in the spectra is $0.6-1.0\arcsec$ along the slit. 
There were patchy cirrus clouds on the first night and photometric conditions 
on the second night.

\begin{figure}
 \figurenum{1}
 \epsscale{1.0}
 \plotfiddle{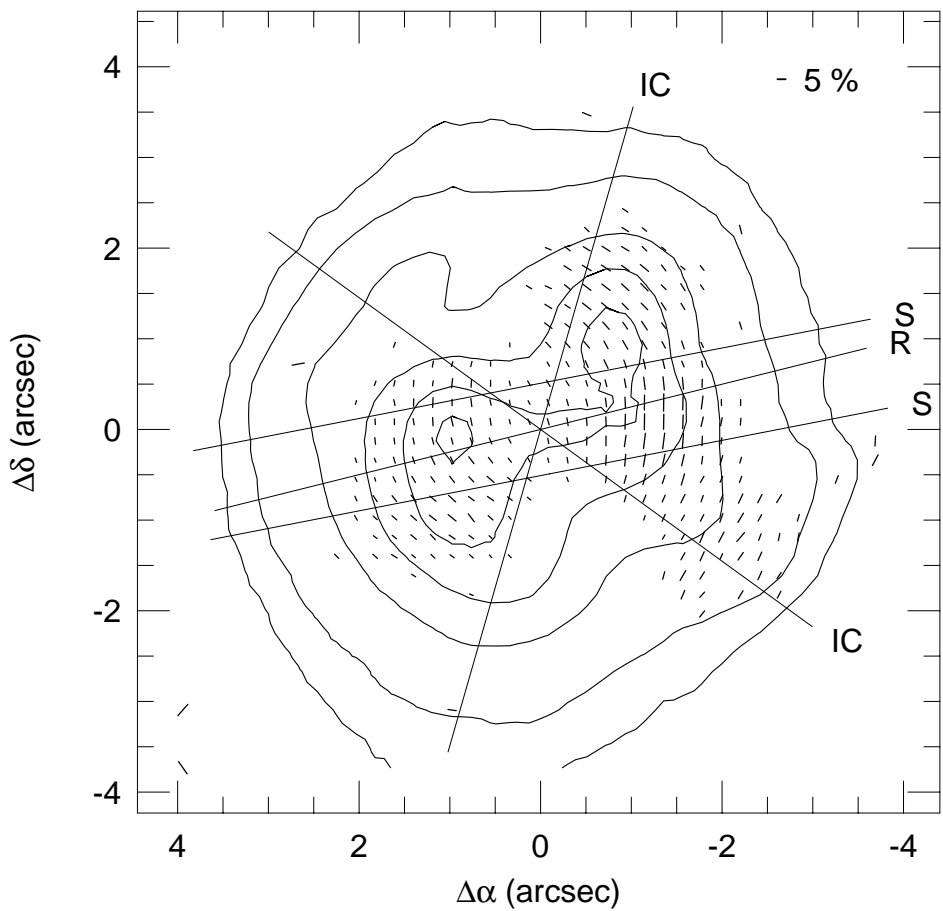}{5in}{0}{70}{70}{-216}{-72}
 \caption[IPOL]{
$B$ band imaging polarimetry of Cygnus A. Polarization vectors with 
SNR$\ge 2.5$ are plotted for each $0.212\arcsec$ pixel over the flux contours 
(20\%, 25\%, 41\%, 60\%, 81\%, 97\%). The centrosymmetric pattern of vectors 
indicates scattering of light from a hidden central source. This scattered 
light cone matches the $55\arcdeg$ half-angle ionization cone (IC) seen in HST
images. The radio axis (R) and slit used for spectropolarimetry (S) are also 
indicated.}
 \label{impol:fig}
\end{figure}

The Galactic interstellar reddening is determined from a  $m_V=19.6$ star 
$10\arcsec$ east of Cygnus A which was also on the slit. We classify this star
as an F9V main sequence star using standard templates (\cite{jhc84}) and 
derive a reddening index of $E(B-V)=0.495\pm0.015$ from the Cardelli, Clayton,
\& Mathis (1989) curve. The corresponding extinction is $A_V=1.5$ magnitudes. 
After correcting for extinction, the spectroscopic distance to the star is 6.0
kpc (610 pc above the Galactic plane), making it a good probe of Galactic 
reddening. Our value for $E(B-V)$ is significantly greater than the average 
(0.36) given by Spinrad \& Stauffer (1982) for galaxies within $4\arcmin$ of 
Cygnus A. However, it is consistent with reddening estimates obtained by 
Tadhunter (1996) from the Cygnus A $H\alpha/H\beta$ line ratio .

We correct our spectra for Galactic reddening, then subtract a scaled template
derived from the elliptical galaxy NGC 821. The best fit for the galaxy 
fraction at 5500~\AA~ is $F_g=70\% \pm 2\%$, obtained using the procedure of 
Tran (1995a). Interstellar Na I D absorption in our Galaxy shows an equivalent
width in the presubtracted spectrum of $2.0\pm0.1$~\AA, broadly consistent 
with the Galactic reddening (\cite{ba86}). We do not correct the data for 
interstellar polarization, which is expected to be small in the $B$ band 
(\cite{jt93}). A correction of 0.5\% would  rotate most of the Cygnus A 
polarization vectors by about $3 \arcdeg$. 

\section{Results}
 
Figure 1 shows the imaging polarimetry of Cygnus A. This $B$ band image 
reveals two fans of scattered polarized light, with the symmetry of the 
vectors clearly indicating a central source. After considering the effects of 
seeing, the scattered light cone matches the ionization cone with $55\arcdeg$ 
half-opening angle  seen in the HST image of Jackson et al. (1996, JTS). The 
outline of this cone is shown in Figure 1, with the apex chosen to be 
consistent ($\pm 0.5\arcsec$) with the polarization pattern and the HST image.
The $B$-band polarization peaks at $10.8\% \pm 0.8\%$ at the position of the 
western knot in the HST image by JTS. The intrinsic polarization of this knot 
must be even greater after correcting for unpolarized galaxy light.

We present the spectropolarimetry in Figures 2 and 3. Figure 2 shows the flux 
and polarization spectra integrated along $7.6\arcsec$ of the slit. $P$ rises 
from $\sim$1\% in the red to 6\% in the blue, and there is a broad rise in $P$ 
centered on H$\alpha$. $P$ drops in the narrow lines, which have $P=1.3\%$ at
$PA=32\arcdeg$. In polarized flux (Fig. 3), the narrow lines are redshifted 
$110-230$ km s$^{-1}$, with the redshift greatest in the high ionization lines
in the eastern cone. The polarized narrow lines are not broadened, suggesting 
that their polarization is at least partly due to dust scattering in an 
outflowing wind, analogous to the phenomenon observed in planetary nebulae 
(\cite{wc94}). However, transmission through dust in our Galaxy and Cygnus A 
may affect the narrow line polarization.

\begin{figure}
 \figurenum{2}
 \epsscale{1.0}
 \begin{center}
 \plotfiddle{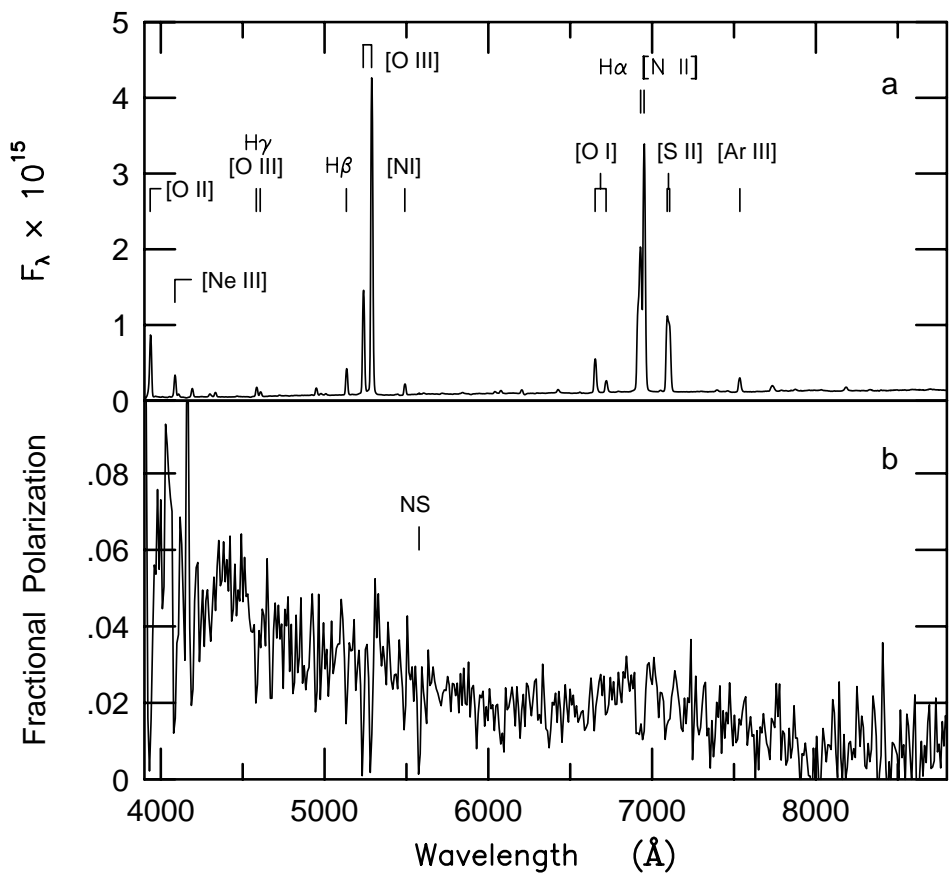}{5in}{0}{70}{70}{-216}{-144}
 \caption[SPOL]{
Spectropolarimetry of Cygnus A. The spectra are integrated along $7.6\arcsec$
of the slit, including  the nucleus, eastern lobe, and western lobe (see Fig. 
1). (a) Total flux. (b) Observed fractional polarization. $P$ rises from 
$\sim$1\% in the red to 6\% in the blue. $P$ drops in the narrow emission 
lines and rises at the location of broad H$\alpha$. The noise feature at 
5577~\AA~ is due to night sky (NS) line emission.}
 \label{spol:fig}
 \end{center}
\end{figure}

Separate spectra of the eastern, western and nuclear components are extracted
using widths of $1.3\arcsec$, $0.8\arcsec$, and $1.1\arcsec$, respectively. 
The galaxy fractions at 5500~\AA ~in the eastern and western components are 
$64\%\pm3\%$ and $62\%\pm3\%$, respectively. Broad lines are visible in the 
dereddened polarized flux spectra (Fig. 3c, d) and in the dereddened, galaxy 
subtracted total flux (Fig. 3a, b) of the eastern and western components. 
Broad lines are also present in the nuclear component at a lower level. Broad 
H$\alpha$ has a Galactic extinction-corrected total flux of $5 \times 10^{-15}
{\rm ~erg ~s}^{-1}{\rm ~cm}^{-2}$ and a full-width at half-maximum (FWHM) of 
$\sim 26,000$ km s$^{-1}$ in both the eastern and western components.  This is 
extremely broad, similar to the broadest radio galaxy emission lines (e.g., 
3C 332, \cite{eh94}). Broad H$\beta$ is also visible in the polarized flux of 
the western component (Fig. 3d). The polarization of the western component 
rises to 16\% at 3800~\AA, suggesting that a large fraction of the continuum 
is scattered light.

\begin{figure}
 \figurenum{3}
 \epsscale{1.0}
 \begin{center}
 \plotfiddle{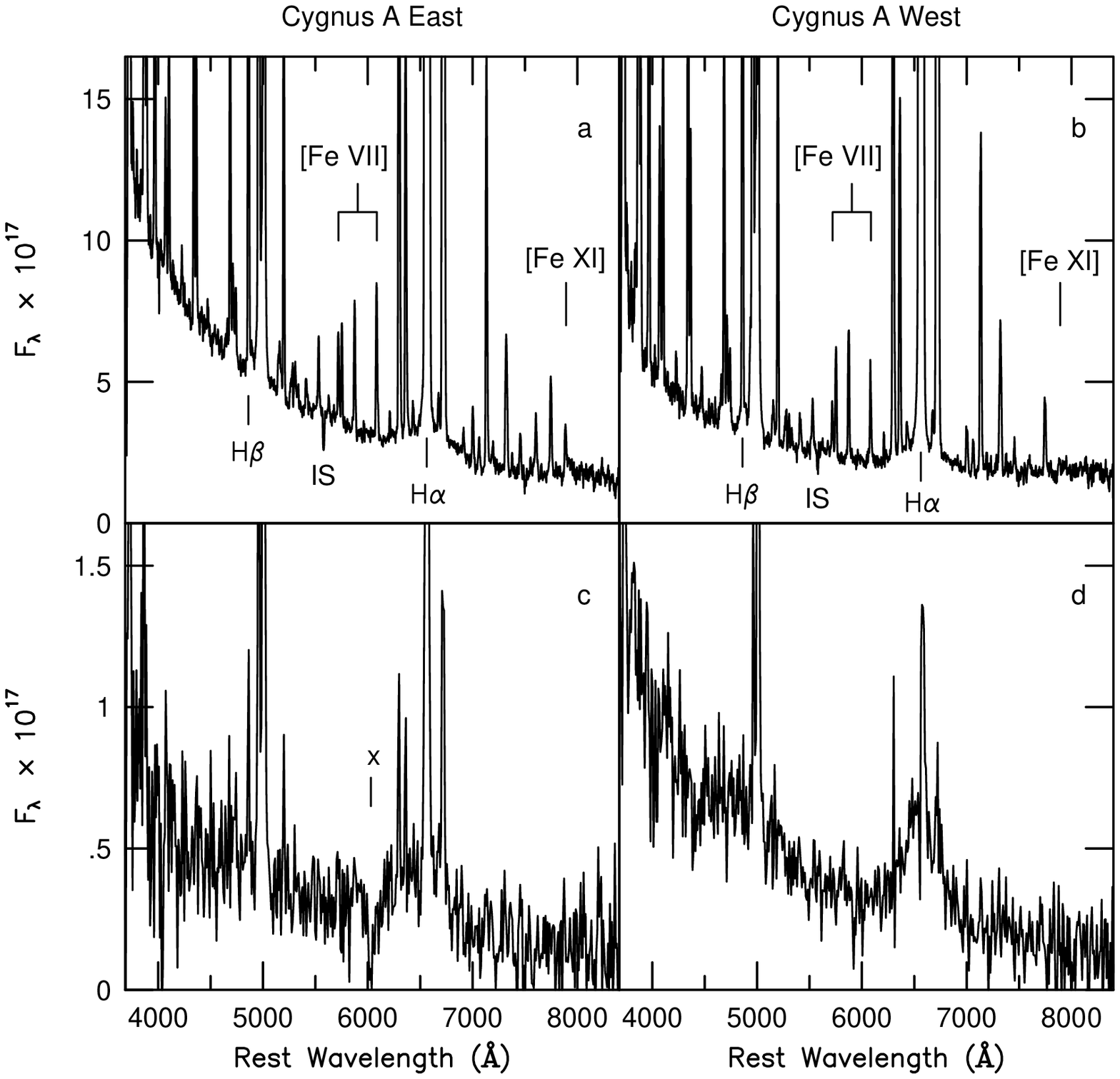}{5in}{0}{80}{80}{-253}{-200}
 \caption[EWSPOL]{
Separate extractions of the eastern and western components of Cygnus
A. (a, b) Galaxy subtracted flux. (c, d) Polarized flux (bin=4). Scattered 
blue continuum and broad H$\alpha$ emission are most prominent in the western 
component. There is an extremely blue contribution to the total flux in the
eastern component which does not show up in polarized flux. The high 
ionization iron lines are relatively stronger by about a factor of two 
in the eastern component. Galactic interstellar Na I D absorption (IS) is 
also apparent. In panel c, the dip at 6000~\AA~ is due to a cosmic ray event.}
 \label{ewspol:fig}
 \end{center}
\end{figure}

Our high quality spectra of Cygnus A display emission lines with a wide range 
of ionization. Identifications are made of [Fe III], [Fe V], [Fe VI], [Fe VII]
and [Fe XI] but the feature at 5305~\AA~ can probably be attributed to [Ca V] 
rather than [Fe XIV], and there is only a marginal detection of [Fe X] 
$\lambda$6375. The most significant changes in relative line intensity with 
position along the slit are for [Fe VII] and [Fe XI], which are stronger to 
the east of the nucleus (Fig. 3). 

\section{Discussion}

\subsection{Hidden AGN}

The symmetric pattern of polarization vectors in the imaging polarimetry 
strongly suggests that we are seeing light from the central AGN scattered 
by material inside a bi-cone of illumination defined by an obscuring torus. 
While Tadhunter et al. (1990) found $P=1\%-2\%$ in $V$ band, perpendicular to 
the radio axis, and concluded that Cygnus A is a ``giant reflection nebula'', 
the much higher degree of polarization (11\%) and greatly improved resolution 
in our $B$-band map make a stronger case for this interpretation. VLBI 
observations of the jet/counterjet flux ratio (\cite{cbr} and references 
therein) constrain the inclination of the jet to $50\arcdeg<i<85\arcdeg$, and 
show that the western side is in front of the eastern side. We obtain an 
independent estimate of the inclination by assuming that the radio and 
ionization cone axes coincide and that the line-of-sight must be outside this 
cone for the nucleus to remain hidden. In this case, the cone half-angle lies 
in the range $46\arcdeg<\theta<55\arcdeg$, consistent with the Barthel (1989) 
hypothesis; and the inclination lies in the range $46\arcdeg<i<90\arcdeg$, 
which matches the VLBI constraints. Forward scattering by dust should increase
the contribution of scattered light in the front cone relative to the 
unpolarized components, thereby raising the fractional polarization in the 
west. 

Seen in polarized flux (Fig. 3), Cygnus A resembles a broad-line AGN . The 
observed ratio of broad H$\alpha$ to broad Mg II (see AHK) in the eastern 
component is 2.0, which becomes 0.38 after correcting for Galactic reddening. 
Typical quasar ratios are 2-3 (\cite{fet91}; \cite{o89}), suggesting that the 
broad lines are scattered by dust with efficiency following a 
$\sim \lambda^{-2}$ law. However, the H$\alpha$ FWHM (in km s$^{-1}$) is 3.5 
times that measured for Mg II by AHK, making the comparison of these lines 
questionable. The broad lines are reflected from extended material at $\sim1 
{\rm~kpc}$, making variability an unlikely explanation of previous H$\alpha$ 
nondetections. Instead, the unusually large width of H$\alpha$ and large 
galaxy fraction probably led to previous underestimates of upper limits to the
H$\alpha$ flux.

\subsection{Spectral Fitting}

The variation of $P$ and $PA$ across the spectrum of Cygnus A can be 
characterized with three differently polarized continuum components, each 
with a wavelength independent $P$ and $PA$ (Fig. 4). There is a nebular 
continuum computed from the observed (Galactic extinction corrected) H$\beta$ 
narrow line flux of $2.2\times 10^{-14} {\rm ~erg ~cm}^{-2} {\rm ~s}^{-1}$. 
This is then reddened by $E(B-V)=0.69$ and given the same $P$ and $PA$ as the 
narrow emission lines. The remaining flux is attributed to a scattered 
continuum, FC1, polarized similarly to broad H$\alpha$ and another featureless
continuum (FC) which is redder and has lower polarization at a different 
$PA$. Note that FC is different from the FC2 component invoked to explain the 
polarization behavior of other AGN (\cite{t95b}; \cite{tcg95}) because it is 
polarized. Figure 4 shows the result of the continuum component fitting to the 
spatially integrated spectropolarimetric data. The blueness and high 
polarization of FC1 is consistent with dust scattering. The redder FC 
component could include locally scattered and/or dichroically absorbed 
radiation from hot young stars associated with the knotty structures seen in 
the HST image (JTS) and analogous to the polarized radiation seen in the dust 
band in Centaurus A (\cite{set96}).

\begin{figure}
 \figurenum{4}
 \epsscale{1.0}
 \begin{center}
 \plotfiddle{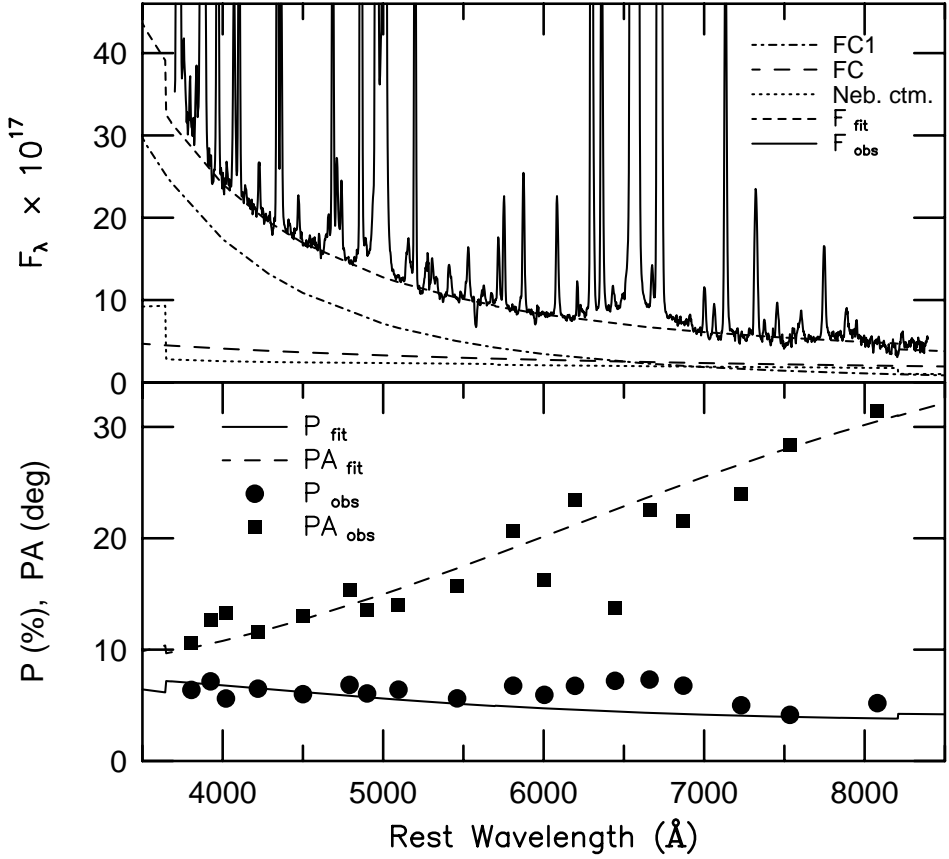}{5in}{0}{70}{70}{-216}{-144}
 \caption[FIT]{
Continuum fits to the spatially integrated spectropolarimetric data.
Observations are corrected for unpolarized host galactic starlight. $P$ and 
$PA$ are calculated in wavelength bins which avoid strong emission lines and 
are therefore, with the exception of several points near H$\alpha$, 
dominated by continuum flux. The nebular continuum has been included with the 
same polarization properties as the narrow emission lines. The combination of 
two additional continua having power-law spectra with different (wavelength 
independent) $P$ and $PA$ is able to well represent the observed continuum 
flux and polarization properties. FC1 has $F_{\nu} \propto \nu^{\alpha}$ with 
$\alpha=+2$, $F_{6563}=2.4\times 10^{-17} {\rm ~erg ~cm}^{-2}{\rm ~s}^{-1}
{\rm ~\AA}^{-1}$, $P=9\%$ and $PA=6^{\arcdeg}$. FC has $\alpha=-1$, 
$F_{6563}=2.5\times 10^{-17} {\rm ~erg ~cm}^{-2}{\rm ~s}^{-1}{\rm~\AA}^{-1}$, 
$P=6\%$ and $PA=48\arcdeg$. Fitting uncertainties in $P$ and $PA$ are 
approximately 1\% and $3\arcdeg$, respectively.}
 \label{fit:fig}
 \end{center}
\end{figure}

The colors of the total and polarized flux vary dramatically between the 
eastern and western components (Fig. 3). The polarized flux is bluer in 
the western component than in the eastern component, but the opposite is true 
of the total flux spectra. The spectral indices (4000~\AA$-$8000~\AA, rest) of
the total flux from the eastern and western components after dereddening and 
galaxy subtraction are $\alpha=0.4$ and $\alpha=-0.4$, respectively, but a 
single power law ($f \sim \nu^{\alpha}$) is not a good fit. Models of the 
spatially separated components will be given in a later paper, but a fit to 
the full data set shows that FC1 dominates in the blue and in the western 
lobe, while the other FC component dominates in the red and in the eastern 
lobe. Furthermore, there must be significant internal reddening of both FC1 
and FC in the western lobe but little in the east.

\subsection{Luminosity of the Hidden AGN}

Cygnus A contains a broad-line AGN, but is it luminous enough to be classified
as a quasar? We estimate the absolute magnitude ($M$) of the central source 
from the highly polarized flux scattered by the western knot. For optically 
thin scattering,

\begin{displaymath}
M = M_{PF} - 2.5 {\rm log} ({1 \over q \tau P}) - A
\end{displaymath}

\noindent
where $M_{PF}$ is the absolute magnitude of the polarized flux, $q$ is the
central source covering fraction, $\tau$ is the scattering optical depth,
$P$ is the intrinsic polarization of the scattered light, and $A$ is the
extinction. We measure $M_{PF}=-14.4$ (rest $B$ band) for the western 
component, assuming a Hubble constant of 73 km s$^{-1}$ Mpc$^{-1}$ and 
correcting for Galactic extinction. The polarization of the scattered light is
at least as great as the measured value in the rest $B$ band, $P=11\%$. The 
source covering fraction of the western knot is estimated to be $q=0.86\%$ 
from the JTS image.  Using these values we find $M_B = -22.0 + 2.5 {\rm log}
~\tau - A_B$.

An optical depth $\tau \le 0.4$ or extinction $A_B \ge 1.0$ would elevate 
Cygnus A to the status of a quasar ($M_B \le -23$), but it would still be
weak compared to other hidden quasars such as 3C 234 and 3C 265 (Tran et al. 
1995; \cite{ctog96}). This agrees with Tadhunter (1996) who concludes that 
Cygnus A contains a ``feeble'' hidden AGN. In addition, the far-infrared 
luminosity of Cygnus A is extremely weak for its radio lobe power (see, e.g., 
\cite{ba96}). So while Cygnus A may qualify as a quasar, it is probably a 
modest one. 

\section{Conclusions}

Cygnus A, the prototypical narrow-line radio galaxy and by far the most 
powerful radio source out to a redshift $z=1$, contains a hidden broad-line 
AGN. The double fan of centrosymmetric polarization vectors seen in imaging 
polarimetry clearly indicates a scattering origin, and corresponds to the 
ionization cone seen in HST images. AGN light is reflected from material 
which has a direct view of the central source. Spectropolarimetry reveals 
broad H$\alpha$ and H$\beta$ emission lines and a blue continuum (FC1) seen 
in scattered light in the eastern and western fans. This confirms the radio 
galaxy-quasar unification by orientation hypothesis. The ratio of broad lines,
blueness of FC1, and narrow line polarization are all consistent with 
scattering by dust. We also find a second polarized continuum (FC), which is 
redder than FC1 and has a different $PA$. The luminosity of the central source
is still uncertain, but appears to be weak for its radio power. 
 
\acknowledgments
 
We thank the Keck Observatory staff who made our observations with the newly 
commissioned Keck II telescope possible. We also thank R. Blandford, S. 
Djorgovski, and J. Walsh for helpful discussions. H. D. T. thanks the 
ST-ECF/European Southern Observatory for its hospitality during his visit. The
W. M. Keck Observatory is operated as a scientific partnership between the 
California Institute of Technology and the University of California; it was 
made possible by the generous financial support of the W. M. Keck Foundation. 
This research has made use of the NASA/IPAC Extragalactic Database (NED).

\end{document}